\documentclass[conference]{IEEEtran}
\IEEEoverridecommandlockouts
\usepackage{cite}
\usepackage{comment}
\usepackage{multirow}
\usepackage{svg}
\usepackage{amsmath,amssymb,amsfonts}
\usepackage{algorithmic}
\usepackage{graphicx}
\usepackage{textcomp}
\usepackage{xcolor}
\def\BibTeX{{\rm B\kern-.05em{\sc i\kern-.025em b}\kern-.08em
    T\kern-.1667em\lower.7ex\hbox{E}\kern-.125emX}}

\begin{document}

\title{BugBlitz-AI: An Intelligent QA Assistant}

\author{\IEEEauthorblockN{Yi Yao, Jun Wang, Yabai Hu, Lifeng Wang, Yi Zhou, \\
Jack Chen, Xuming Gai, Zhenming Wang, Wenjun Liu}
\IEEEauthorblockA{Intel \\
\textit{\{yi.a.yao, jun.j.wang, yabai.hu, lifeng.a.wang, yi.a.zhou,
jack.z.chen, xuming.gai,zhenming.wang, wenjun.liu\}@intel.com}}
}

\maketitle

\begin{abstract}
The evolution of software testing from manual to automated methods has significantly influenced quality assurance (QA) practices. However, challenges persist in post-execution phases, particularly in result analysis and reporting. Traditional post-execution validation phases require manual intervention for result analysis and report generation, leading to inefficiencies and potential development cycle delays. This paper introduces BugBlitz-AI, an AI-powered validation toolkit designed to enhance end-to-end test automation by automating result analysis and bug reporting processes. BugBlitz-AI leverages recent advancements in artificial intelligence to reduce the time-intensive tasks of manual result analysis and report generation, allowing QA teams to focus more on crucial aspects of product quality. By adopting BugBlitz-AI, organizations can advance automated testing practices and integrate AI into QA processes, ensuring higher product quality and faster time-to-market. The paper outlines BugBlitz-AI's architecture, discusses related work, details its quality enhancement strategies, and presents results demonstrating its effectiveness in real-world scenarios.
\end{abstract}

\begin{IEEEkeywords}
software quality assurance, large language model (LLM), test automation, log analysis
\end{IEEEkeywords}

\section{Introduction}
The evolution from manual to automated testing has significantly shaped quality assurance (QA) practices over the past decades. Despite these advances, the domain of test automation faces persistent challenges, particularly in post-execution phases such as result analysis and reporting. Traditional automated systems efficiently execute tests but often require manual intervention to analyze outcomes and generate reports, such as Jira entries, leading to inefficiencies and potential delays in development cycles.
For post-execution phase of quality assurance, validation engineers have to review all failure logs and identify the root cause of each failure. Then they need to diagnose whether the failures are the quality issues of the product and identify the specific error type of each failure manually. In cases of numerous failures, they also need to determine if any are duplicates. Once all error logs have been analyzed, validation engineers need to summarize the descriptions of each error and report them, often by creating Jira tickets. Despite being critical, these tasks can be monotonous and time-consuming.
This paper introduces BugBlitz-AI, an innovative AI-powered validation toolkit designed to enhance end-to-end test automation by automating the analysis of test results and bug reporting processes. By leveraging recent advancements in artificial intelligence, BugBlitz-AI aims to reduce the time-intensive tasks of manual result analysis and report generation, thus allowing QA teams to concentrate more effectively on crucial aspects of product quality.
The adoption of BugBlitz-AI not only promises to advance the current state of automated testing but also sets a reference for integrating AI into QA processes, thereby ensuring higher product quality and faster time-to-market. This research highlights the potential of AI to transform key aspects of QA, positioning intelligent automation as a critical tool in the software development lifecycle.
The paper is organized as follows. In section 2, we give an overview of related works. In section 3, we introduce the architecture of BugBlitz-AI. In section 4, we discuss the how to enhance the quality of BugBlitz-AI. Section 5, we present the results.

\section{Related Works}

The advancements in artificial intelligence and machine learning are widely employed in automating validation tasks. And for post-execution validation automation, BugBlitz-AI provides 4 LLM sub-modules, root error analysis, bug diagnosis, bug summarization and duplicate detection.

\paragraph*{Root error analysis} it aims to identify the fundamental reason behind multiple errors. Root error analysis is critical to improve the efficiency of issue profiling and further bug fixes. The industry has already begun leveraging machine learning\cite{1_LOKRANTZ20181057} and big data \cite{2_MA2021107580} to conduct root cause analyses of failures.
\paragraph*{Bug diagnosis} it is the task to classify an issue into bug or non-bug categories, effective automatic classification can significantly save manual efforts and improve efficiency in development cycles. Several research have proposed machine learning based solutions for this bug/non-bug binary classification with different designs of textual fields selections, feature representation methods, and ML algorithms\cite{3_li2024empirically}. More advanced studies introduce multi-classification solutions to further categorize different types of bugs with ensemble machine learning approach with text augmentation technique \cite{4_alsaedi2023nature}, stacking classifier \cite{5_kusumaniswar2024bug}, etc.
\paragraph*{Bug summarization} it involves condensing information such as failed test case, code snippets and error messages etc. into concise and informative bug summary and description. To explore and enhance the effectiveness of bug summarization, researchers in both industry and academia have proposed various methods including deep learning utilizing the sentence importance \cite{6_YKoh2022deeplearning}, deep attention-based approach using the pretrained RoBERTa encoder \cite{7_Ma2023attention} , NLP technique \cite{8_Irtaza2020nlp}, ML sentences extraction using classifiers on diverse features \cite{9_Rastkar2014automatic}.
\paragraph*{Duplicated Bug Report Detection (DBRD)} it poses a persistent challenge in both academic research and industrial practice. The advent of deep learning has spurred the emergence of numerous approaches aimed at addressing this issue in recent years \cite{10_8094414} \cite{11_3389263} \cite{12_3387470} \cite{13_6100061} \cite{14_9251082} \cite{15_zhang2023cupid}.

Although various AI-driven research and engineering initiatives have been undertaken for specific log analytics tasks as mentioned above, a comprehensive pipeline that fully automates the bug analysis and reporting process is still lacking. BugBlitz-AI steps in to fill this void.

\section{BugBlitz-AI}
\begin{figure}[htbp]
\centerline{\includegraphics{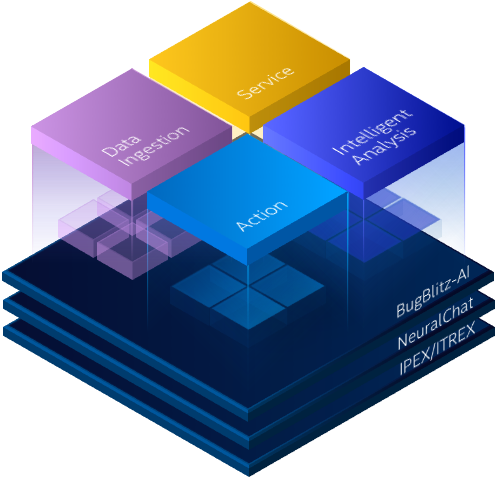}}
\caption{BugBlitz-AI Architecture}
\label{arch}
\end{figure}

Figure \ref{arch} shows the architecture of the BugBlitz-AI. It consists of four key modules: Service, Data Ingestion, Intelligent Analysis, and Action. The core component of BugBlitz-AI is Intelligent Analysis module, which is powered by generative AI. It has 4 Large Language Model (LLM) submodules. The software stack BugBlitz-AI used is NeuralChat \cite{neuralchat}, Intel\textsuperscript{@} Extension for PyTorch \cite{ipex} and Intel\textsuperscript{@} Extension for Transformers \cite{itrex}.

\subsection{Service Module}\label{AA}
Service module plays a central role in the BugBlitz-AI architecture. BugBlitz-AI offers a standalone service that provides Python and RESTful APIs, allowing seamless integration with all existing automatic validation pipelines. The service module receives requests from the validation pipelines, then triggers the data ingestion module for data retrieval. Subsequently, it forwards the output of the data ingestion module to the Intelligent Analysis module. Finally, depending on the output of the Intelligent Analysis module, the service module will call the action module to conduct the appropriate action.

\subsection{Data Ingestion Module}
The input of BugBlitz-AI is the detailed information of the failed test cases and data ingestion module is responsible for handling these data. The information usually includes two parts, error log and test information. For the error log, the data ingestion module matches errors from the error logs with a customized error pattern list. For test information, the data ingestion module integrates this information and passes them over to the service module.

\subsection{Action Module}
Action module provides the specific implementation for the post-execution work. Now BugBlitz-AI supports two actions: posting Jira ticket and sending email notification. The choice of action is determined by the Intelligent Analysis Module’s assessment.

\subsection{Intelligent Analysis Module}

After the raw log is processed by data ingestion module, the extracted error information is subsequently managed by intelligent analysis module, powered by generative AI. The intelligent analysis module is composed of four sub-modules, next we will introduce the roles of each sub-module and the logical relationship among them: 
\begin{enumerate}
    \item Root Error Analysis sub-module: it is designed to identify the root cause of multiple errors raised from a single test failure. 
    \item Bug Diagnosis sub-module: it receives the root cause error identified by root error analysis module, and then determines whether the error is caused by an actual bug in the code, or by non-bug issue such as misconfigured test environments. If the error is judged to be a bug, the error information will be passed on to the subsequent sub-modules.
    \item Bug Summarization sub-module: it generates the summary of the bug in the format required by the bug reporting system (e.g., Jira system).
    \item Duplicate Detection sub-module: it takes in multiple error summaries generated by bug summarization sub-module, and determines whether these summaries correspond to duplicated errors. It addresses the scenario where multiple test cases are submitted in one request, among which several cases are caused by the same bug; or the current bug had been reported but not yet fixed. In these cases, if these bugs are determined as duplicates, redundant bug reports can be avoided.
\end{enumerate}

\begin{figure*}[htbp]
\centering
\includegraphics[width=0.85\textwidth]{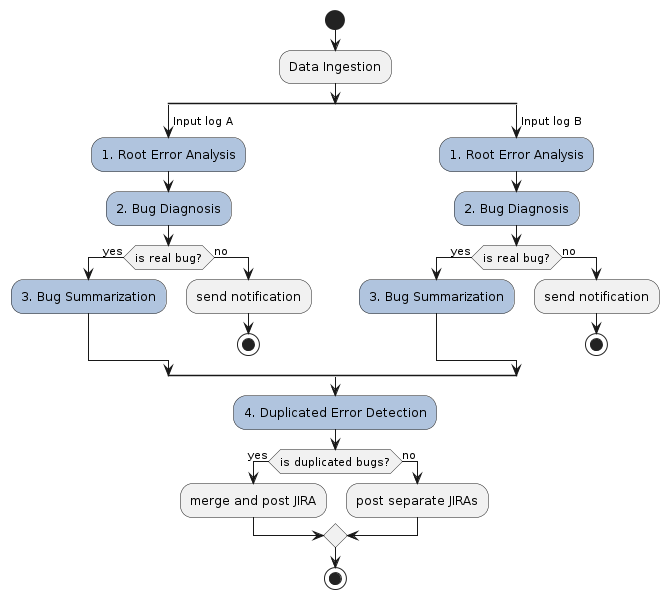}
\caption{BugBlitz-AI Workflow}
\label{workflow}
\end{figure*}

\textit{Model Selection:}
The intelligent analysis sub-modules in BugBlitz-AI are implemented by feeding large language models with appropriate information and instructions. Given the constraints of computation resources, we do not anticipate employing a single large-scale model for multiple tasks. Instead, our approach involves utilizing smaller, faster models tailored to specific tasks individually. Hence, we initiate with a relatively small model size at around 7B, and adopt different instruction models for each sub-module according to the required capabilities of their corresponding task.
For root error analysis sub-module, it is responsible for identifying the primary root cause error among a possibly deep stack consisting of multiple errors, the task requires a higher capability of log comprehension and code analysis, so we employed a state-of-the-art (as the date of 2024.2, \cite{17_bigcode}) instruction model finetuned on coding data of our required model size: DeepSeek-Coder-7b-instruct \cite{16_deepseek-coder}. For bug diagnosis sub-module, its task is to determine whether a failure is caused by a real bug based on the error information and judging criteria given in the prompt, so it requires higher context comprehension ability rather than code analysis ability. Therefore, we chose a more general and comprehensive model: Mistral-7B-Instruct \cite{18_jiang2023mistral}. For bug summarization and duplicate detection sub-modules, we used CodeLlama-7b-Instruct \cite{19_rozière2024codellama} model. 
The table \ref{models} summarizes the selected models used in BugBlitz-AI. 

\begin{table*}[hbtp]
\centering
\begin{tabular}{|l|l|l|l|l|}
\hline
           & \textbf{Root Error Analysis}        & \textbf{Bug Diagnosis}       & \textbf{Bug Summarization}     & \textbf{Duplicated Error Detection} \\ \hline
LLM models & DeepSeek-Coder-7b-instruct & Mistral-7B-Instruct & CodeLlama-7b-Instruct & CodeLlama-7b-Instruct      \\ \hline
\end{tabular}
\caption{Models used in intelligent analysis sub-modules}
\label{models}
\end{table*}

The workflow of the Intelligent Analysis module is illustrated in Figure \ref{workflow}.

\section{Quality Enhancement }

BugBlitz-AI leverages the LLM technology to implement the automation for post-execution of validation work. In this section, we discuss the quality enhancement of BugBlitz-AI by task decoupling, prompt engineering and fine-tuning.

For the quality, recall and precision serve as pivotal performance metrics in an error analysis and reporting system. Recall is the fraction of bugs that were identified by BugBlitz-AI:
\begin{equation}
Recall=\frac{\text{bugs identified by BugBlitz-AI}}{\text{all bugs}}
\label{recall}
\end{equation}
Precision measures the percentage of BugBlitz-AI filed Jira tickets that necessitate no human intervention:
\begin{equation}
Precision=\frac{\text{posted Jira tickets without human intervention}}{\text{posted Jira tickets}}
\label{precision}
\end{equation}

\subsection{Improving Recall}

Recall represents the percentage of the bugs accurately identified by BugBlitz-AI: a higher recall equates to lower rate of missed bug reports. Therefore, maintaining recall at a high standard is crucial to ensure the quality of BugBlitz-AI. 
Bug diagnosis sub-module is the primary component that can influence recall. If it misclassifies an actual bug as non-bug, it will negatively impact recall.
To mitigate such misjudgment of bug diagnosis sub-module, we have employed strategies including task decoupling and prompt engineering. 
\subsubsection{Task Decoupling}
In the initial design of bug diagnosis task, we fed the model with both the complete error list and the task description. However, this approach didn’t produce satisfactory results because it implicitly incorporated an additional task: root error analysis. This added complexity is proved to be too challenging for the LLM to handle effectively. Therefore, we decouple these two tasks into distinct sub-modules: our current root error analysis and bug diagnosis sub-module, and provide only the root cause error to bug diagnosis sub-module. This separation of tasks allows the bug diagnosis sub-module to be solely responsible for classifying a single error, contributing to a substantial improvement in the output quality.
\subsubsection{Prompt engineering}
We also utilized methods of prompt engineering to further improve the generation quality of bug diagnosis sub-module, including few-shot prompt and chain-of-thought:
\begin{itemize}
\item Few-shot Prompt: When giving the task description cannot produce satisfactory result, we can leverage the in-context learning capability of the LLM by directly providing examples of question-and-answer that demonstrates our expected response. For example, in the root error analysis sub-module, we expect the LLM to identify the root cause error from the given error list and return its error number. Providing an output example, rather than a task description, effectively guides the LLM to generate the required response, particularly for restricting the output format for the root error analysis sub-module.
\item Chain of Thought \cite{20_wei2023chainofthought}: When applying few-shot prompt technique, for some complex tasks, simple examples of question-answer pair are not sufficient for the LLM to learn the internal logic that maps the input to the desired output. In this circumstance, we can provide intermediate steps that help the model in reasoning towards the correct result. In the bug diagnosis sub-module, the task of the LLM is to determine if an error is an actual bug based on our provided judging criteria. When using few-shot prompt, it becomes necessary to combine with chain of thought technique that provides a reasoning step that demonstrates the logic behind making the correct judgment with the given criteria.
\end{itemize}
Table \ref{bug_diagnosis_prompt} shows a prompt example of bug diagnosis sub-module integrated with above techniques.

\begin{table}[htbp]
\begin{tabular}{c}
\includegraphics[width=\linewidth]{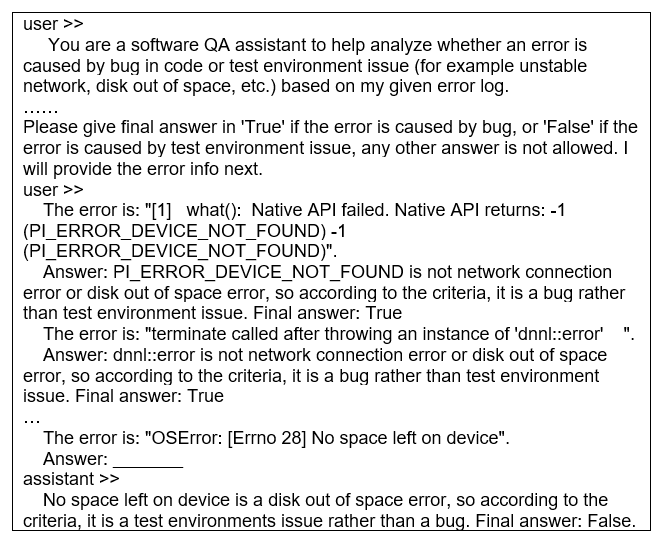} \\
\end{tabular}
\caption{Bug Diagnosis Prompt Example}
\label{bug_diagnosis_prompt}
\end{table}

\subsubsection{Quality Evaluation Results}
\begin{figure}[htbp]
\centering
\includegraphics[width=\linewidth]{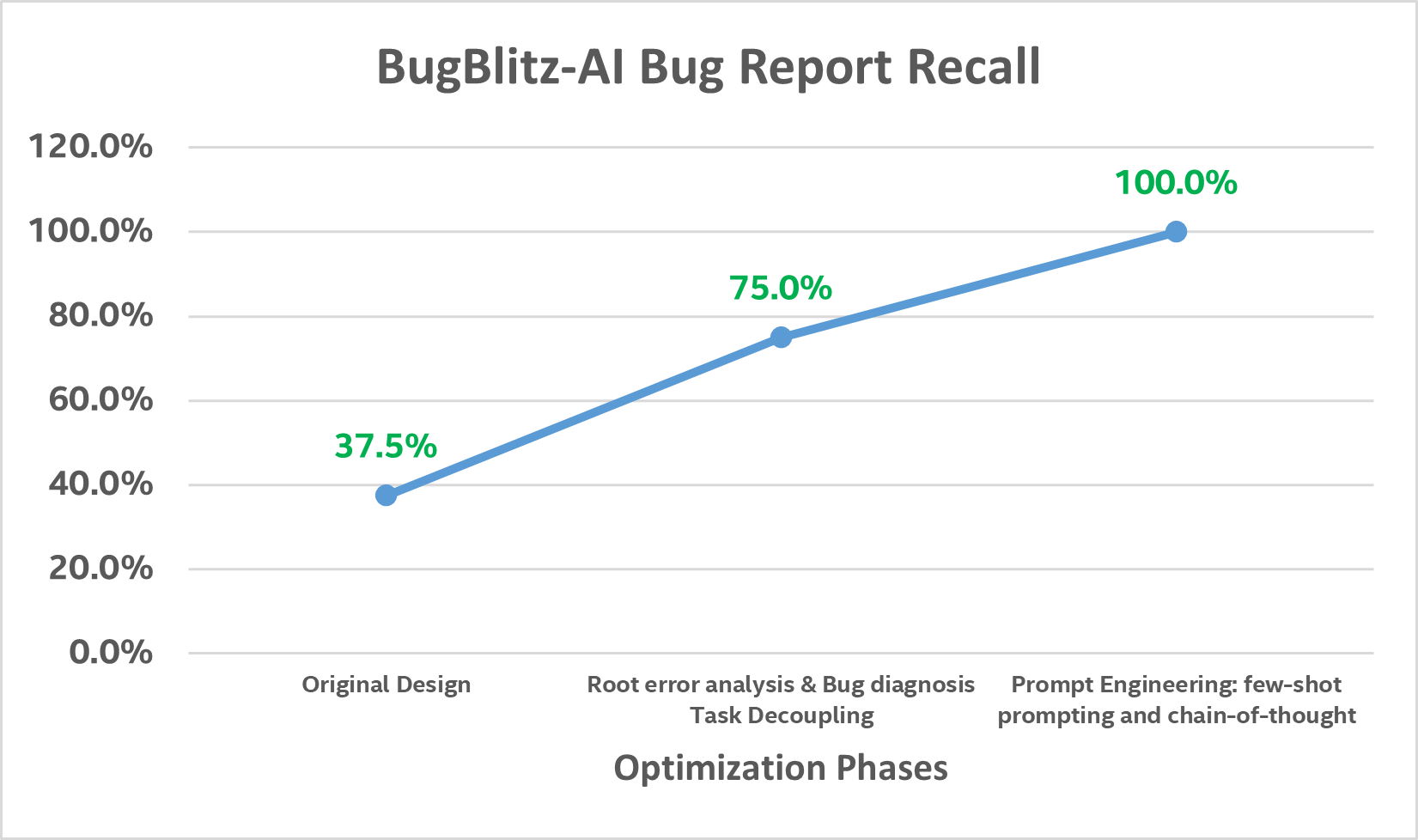}
\caption{BugBlitz-AI Bug Report Recall}
\label{recall-fig}
\end{figure}

By applying the above optimization strategies, the recall of the BugBlitz-AI can be considerably enhanced. We have evaluated the recall on a dataset collected from 1k bugs extracted from software products in a real production environment. Evaluation results in figure \ref{recall-fig} illustrates the recall improvement brought by the above optimizations.

\subsection{Improving Precision}
Precision represents the percentage of the bug reported by BugBlitz-AI that can be accepted without human intervention. A higher precision signifies a greater efficiency improvement achieved by BugBlitz-AI. We have adopted both prompt engineering and fine-tuning techniques to improve precision.
\subsubsection{Prompt engineering}
To improve the precision of the bug summarization sub-module with a more effective prompt, we adopted prompt chaining technique:
\begin{itemize}
    \item Prompt Chaining: This is a primary technique to improve the performance of LLM. It involves decomposing the original complex task into multiple sub-tasks, each with its distinct prompt. These prompts are combined into a prompt chain guiding the model through the completion of the complex task. In the bug summarization sub-module, we need the model to generate a formatted summary and description of a bug based on the given error information. If we instruct the model to complete this task in one prompt, the model generally gives unsatisfactory result. So we break the instruction of the original task into three prompts: 1. State the task’s general purpose; 2. Provide the actual error information; 3. Instruct to model to refine the original answer according to the given output format. By dividing the task instruction into a prompt chain, we observe an obvious improvement of the quality of the generated content.
\end{itemize}
Table \ref{bug_summ_prompt} shows a prompt example of bug summarization sub-module with prompt chaining:

\begin{table}[htbp]
\begin{tabular}{c}
\includegraphics[width=\linewidth]{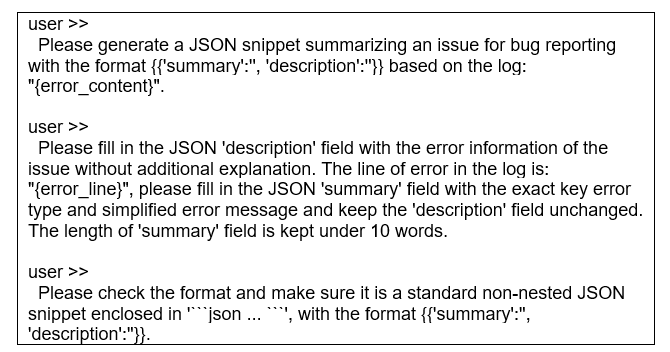} \\
\end{tabular}
\caption{Bug Summarization Prompt Example}
\label{bug_summ_prompt}
\end{table}

\subsubsection{Fine-tuning}
In BugBlitz-AI, the precision relies on the performance of the LLMs used on the four intelligent analysis sub-modules. While many LLM base models are designed for general purposes, our utilization of fine-tuned open models tailored for programming tasks has not yet yielded the desired levels of precision and recall. So, fine-tuning is required to adapt pre-trained LLM base models to error analysis tasks.
We take the following steps to fine-tune BugBlitz-AI models: a). Prepare training data; b). Training a new fine-tuned model; c). Hyperparameter optimization.
\newcommand{\subsubsubsection}[1]{\paragraph{#1}\mbox{}\\}
\setcounter{secnumdepth}{4}
\setcounter{tocdepth}{4}
\paragraph{Preparing training data}
It’s a significant challenge to create a high quality, diversity dataset for LLM training, even if just for specific tasks. Here are some key challenges of creating an ideal dataset for error analysis:
\begin{itemize}
    \item \textbf{Domain Relevance:} The quality of finetuned models heavily depends on the relevance of the data to error analysis domain. There’re various test types, such as unit tests and integration tests etc., each exhibiting distinct log patterns. Collecting data to cover various test types is essential.
    \item \textbf{Data Diversity:} A diverse dataset is crucial for model quality. The dataset needs to cover different scenarios. E.g. The test logs should encompass a wide range of errors, including those originating from the system under test (e.g., users' software products, software dependencies, drivers, operating systems) as well as test environment issues (e.g., network timeouts, disk space exhaustion).
    \item \textbf{Data Labeling:} High-quality labeled data entails concise bug summaries along with the root cause of errors leading to test failures.
    \item \textbf{Data Cleansing:} The raw data we've gathered is rife with errors, noise, inconsistencies, and duplications. For instance, the bug summary label often includes users’ project names that are absent in the input logs. This mismatch poses a significant challenge, as language model models may attempt to infer project names from the training dataset, resulting in erroneous bug titles. Therefore, it's imperative that we prioritize the use of clean data to ensure our models learn accurately and effectively.

\end{itemize}
We encountered a scarcity of open datasets suitable for our error analysis needs, prompting us to take matters into our own hands. To address this gap, we painstakingly crafted a dataset from scratch. This involved extracting raw data directly from the bug tracking systems of several ongoing software projects. Leveraging Jira attached logs as our primary data inputs, we labeled data with human-written Jira tickets. Below is a label example of training samples.

\begin{table}[htbp]
\begin{tabular}{c}
\includegraphics[width=\linewidth]{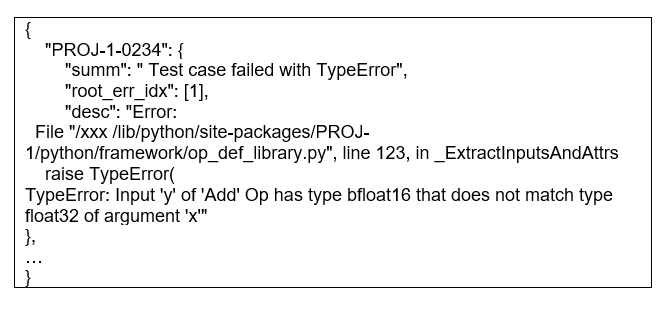} \\
\end{tabular}
\caption{Label Example for Fine-tuning}
\end{table}

\paragraph{Training a new fine-tuned model}
In natural language processing, a pivotal paradigm involves extensive pre-training on vast general domain datasets, followed by adaptation to specific tasks or domains. However, as we scale up our pre-training efforts to accommodate larger models, the practice of full fine-tuning, which involves retraining all model parameters, becomes increasingly impractical. Despite leveraging 7b models within BugBlitz-AI, the prospect of full fine-tuning remains prohibitively costly and challenging to iterate swiftly, especially to accommodate the evolving needs of users' tests. We used Low-Rank Adaption, abbreviated as LoRA \cite{21_hu2021lora}, as a solution. LoRA operates by immobilizing the pre-trained model weights and introducing trainable rank decomposition matrices into every layer of the Transformer architecture. This process significantly diminishes the number of trainable parameters for subsequent tasks. Leveraging the Hugging Face Parameter Efficient Fine-Tuning (PEFT) library \cite{22_xu2023parameterefficient}, LoRA guarantees ease of implementation. 

\paragraph{Hyperparameter optimization}
When employing PEFT for training a model with LoRA, the hyperparameters governing the low-rank adaptation process can be specified within a LoRA config, as illustrated below:

\begin{table}[htbp]
\begin{tabular}{c}
\includegraphics[width=\linewidth]{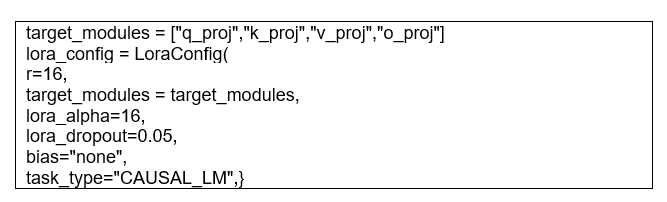} \\
\end{tabular}
\caption{LORA Config Used in Fine-tuning}
\end{table}

Two of these hyperparameters, namely $r$ and $target\_modules$ have been empirically shown to significantly impact adaptation quality and will be the focal points of the forthcoming tests. The remaining hyperparameters will be held constant at the values previously specified for simplicity.
The parameter $r$ signifies the rank of the low-rank matrices learned during the fine-tuning process. As this value escalates, the number of parameters requiring updates during low-rank adaptation also increases. In essence, a lower $r$ might facilitate a quicker, less computationally intensive training process but could potentially compromise the resultant model's quality. However, pushing $r$ beyond a certain threshold may not yield any discernible improvement in the output model's quality. The impact of $r$ on adaptation (fine-tuning) quality will be subjected to testing shortly.
During fine-tuning with LoRA, it becomes feasible to target specific modules within the model architecture. The adaptation process will concentrate on these modules and apply the update matrices accordingly. Similar to the scenario with  $r$, targeting a greater number of modules during LoRA adaptation leads to prolonged training times and increased demand for computational resources. Hence, it's customary to solely target the attention blocks of the transformer. 
\subsubsection{Quality Evaluation Results}
\begin{figure}[htbp]
\centering
\includegraphics[width=\linewidth]{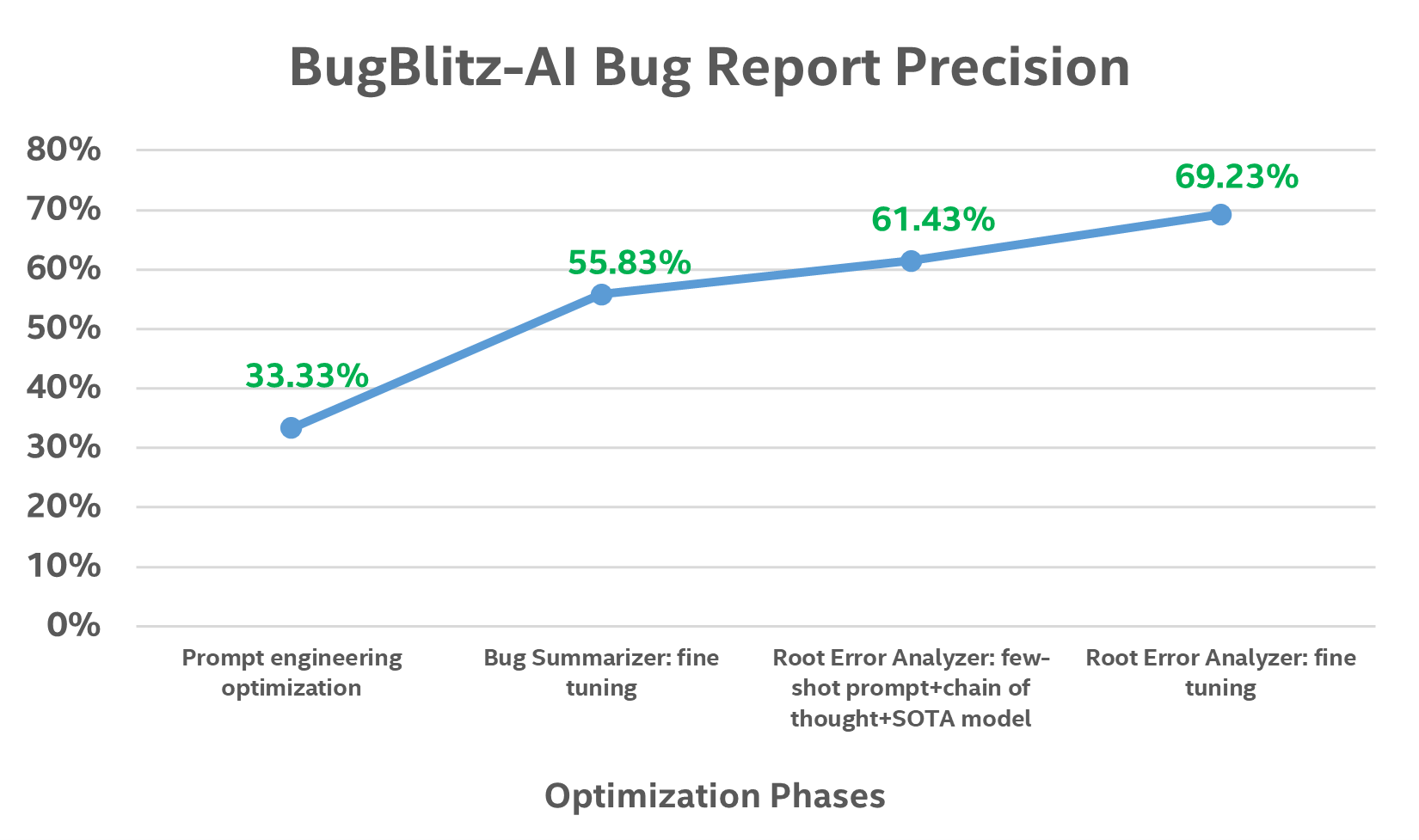}
\caption{BugBlitz-AI Bug Report Precision}
\label{precision-fig}
\end{figure}

We have evaluated the precision on the same dataset used in recall evaluation. The evaluation result in figure \ref{precision-fig} shows the precision improved by the above quality enhancement techniques.

\section{Results}
BugBlitz-AI deployed on Intel\textsuperscript{@} Xeon\textsuperscript{@} Scalable Processors provides an automating solution for validation post-execution by utilizing generative AI and shows 69.3\% precision and 100\% recall in real use scenarios.
BugBlitz-AI significantly streamlines the process, greatly reducing the effort required for QA validation. It enhances the efficiency of validation and the project cycle compared to manual methods.

\bibliographystyle{IEEEtran}
\bibliography{refs} 

\vspace{12pt}

\end{document}